\documentclass[prl, twocolumn, amsmath,amssymb,superscriptaddress]{revtex4}

\usepackage{hyperref}
\usepackage{graphicx}
\usepackage{dcolumn}
\usepackage{xcolor}
\usepackage{braket}
\usepackage{commath}
\usepackage{adjustbox}
\usepackage{comment} 
\usepackage{ulem}
\usepackage{tikz}
\usetikzlibrary{shapes}

\newcommand{\bs}[1]{{\boldsymbol{#1}}}
\newcommand{\bk}{\bs{k}}

\begin{document}

\title{Robust coherent transport of light in multi-level hot atomic vapors}

\author{N. Cherroret}
\affiliation{Laboratoire Kastler Brossel, UPMC-Sorbonne Universit\'es, CNRS, ENS-PSL Research University, Coll\`ege de France; 4 Place Jussieu, 75005 Paris, France}
\author{M. Hemmerling}
\affiliation{Instituto de F\'{i}sica de S\~ao Carlos, Universidade de S\~ao Paulo, 13560-970 S\~ao Carlos, SP, Brazil}
\affiliation{Universit\'e C\^ote d'Azur, CNRS, Institut de Physique de Nice, Valbonne F-06560, France}
\author{V. Nador}
\affiliation{Universit\'e C\^ote d'Azur, CNRS, Institut de Physique de Nice, Valbonne F-06560, France}
\author{J.T.M. Walraven}
\affiliation{Van der Waals-Zeeman Institute, Institute of Physics, University of Amsterdam, Science Park 904, 1098 XH Amsterdam, The Netherlands}
\author{R. Kaiser}
\affiliation{Universit\'e C\^ote d'Azur, CNRS, Institut de Physique de Nice, Valbonne F-06560, France}


\begin{abstract}
Using a model system, we demonstrate both experimentally and theoretically that coherent scattering of light  can be robust in hot atomic vapors despite a significant Doppler effect. By operating in a linear regime of far-detuned light scattering, we also unveil the emergence of interference triggered by inelastic Stokes and anti-Stokes transitions involving the atomic hyperfine structure.
\end{abstract}


\maketitle

Wave propagation in disordered media is at the focus of intense investigations in many different fields of research, including condensed matter \cite{Akkermans07}, astrophysics \cite{Hulst12}, acoustics \cite{Tiggelen08}, optics \cite{Gigan17}, atomic physics \cite{Molisch1998} and ultracold atoms \cite{Aspect09}. Often, many parameters of the system under study are not fully controlled and are described by effective parameters, including for instance the strength of disorder or decoherence. 
Similarly, the internal structure of atoms is often put aside and a simplified two-level model is used to describe the qualitative behavior of observed phenomena. 
Interestingly, in cases when the detailed structure is taken into
account, not only the quantitative description can be
improved, but new qualitative features can emerge, such as Sisyphus cooling \cite {Dalibard1989}, slow light \cite{Hau1999} and quantum memories \cite{DLCZ}. 
Doppler broadening in room temperature vapors adds to the complexity of a microscopic description of coherent light propagation and a radiative transfer equation is often used as a simplified model \cite{Molisch1998}. This however prevents the description of coherence phenomena in nearby coupled dipoles \cite{Guerin2016, Pellegrino14, Browaeys2016, Beugnon2017,Saint-Jalm2018} or the anomaly of polarization at the D1 line of the solar spectrum \cite{Stenflo2015}. Even though Doppler-free spectroscopy in room temperature atomic vapors has seen important results, the development of laser cooling of atoms allowed for an impressive improvement of high-precision measurements.
Presently, we witness a renewed interest in precision measurements with hot atoms,
including applications on electric-field sensors \cite{Shaffer2013} and quantum information science \cite{Pfau2015, Hughes2017, Polzik2017, Mitchell2018}.
In this Letter we report on a coherent light scattering experiment with a hot atomic vapor in which thermal decoherence is largely circumvented. We also show that the
internal multi-level structure of the atoms gives rise to qualitatively new interference features which emerge as a result of inelastic scattering and can be described with a
quantitatively accurate microscopic theory.

\begin{figure}[h]
\centering
\includegraphics[scale=0.8]{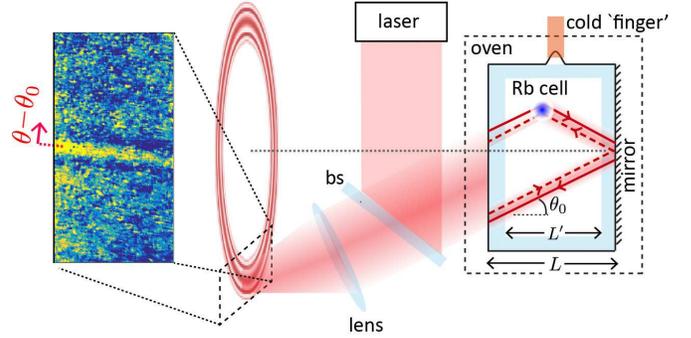}
\caption{(Color online) 
Scheme of the experiment (for clarity polarization elements are not shown). A collimated laser beam is sent via a beamspitter (bs) through a glass cell (width $L\simeq 8$ mm, chamber thickness $L'\simeq6$ mm) containing a hot mixture of rubidium atoms. An oven with a cold `finger' is used to regulate the vapor pressure in the cell. The backscattered signal  is collected in the far field on a CCD camera. 
In the cell, the interference between counter-propagating wave paths scattered on both an atom and the back face of the cell gives rise to annular fringes (``mirror-assisted coherent backscattering'').
\label{Exp_scheme}}
\end{figure}
\begin{figure}
\centering
\includegraphics[scale=0.42]{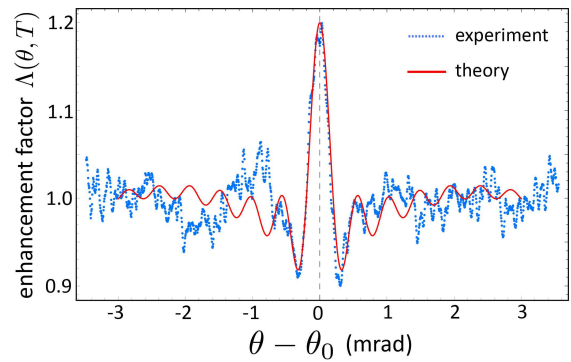}
\caption{(Color online)  Angular radial cut of the mCBS interference ring (channel $h\perp h$, detuning $|\Delta|=2$ GHz, incident angle $\theta_0\simeq 7^\circ$). Blue dots are the experimental signal. The red curve is Eq. (\ref{EFtheta_eq}) in which the amplitude at $\theta=\theta_0$, $\Lambda(\theta_0,T)$, is set to the experimental value 1.20. Except for this reference point, there is no adjustable parameter.
\label{mCBS_width}}
\end{figure}

The scheme of our experiment is shown in Fig. \ref{Exp_scheme}. A collimated laser beam (waist $w = 10$ mm) is sent through a slab-shaped glass cell containing a natural mixture of rubidium vapor at an oven-regulated temperature/density. The wavelength $\lambda=780$ nm is set to the D2 transition of rubidium. Two different cells were used, one with a metallic mirror clipped to the back side, and another one without such a mirror.
The angle of incidence of the laser beam, $\theta_0\ll 1$, was adjusted to typically a few degrees with respect to the surface normal of the slab. The backscattered light was observed in the far field with an optical angular resolution of 0.044 mrad using a CCD camera (Fig. \ref{Exp_scheme}).
As a result of light scattering, this signal has generically two contributions: an ``incoherent'' one,  associated with pairs of optical paths traveling  along a common scattering trajectory in the same direction, and a ``coherent'' one where the two paths propagate along the same trajectory but in opposite directions \cite{Sheng95}. In the multiple scattering regime the latter leads in particular to the coherent backscattering effect, which was previously measured in cold atomic gases \cite{Labeyrie99}. The goal of our experiment is to investigate the coherent component of the signal in a hot vapor. For this purpose, the cell is heated to temperatures on the order of $200^\circ$C. In such a regime, the atomic motion is very fast which a priori constitutes an unfavorable case for coherent transport.
Indeed, the Doppler effect associated with thermal motion  induces a random frequency shift in scattered wave paths, a phase-breaking mechanism usually detrimental to interference \cite{Golubentsev84, Labeyrie06}. 
To counteract this mechanism, our strategy has been to operate at large detuning $|\Delta\equiv\omega-\omega_0|\gg k\overline{v}$ so that the characteristic time $\sim\lambda/\overline{v}$ for an atom to move over a wavelength becomes much longer than the time $\tau\sim\Gamma^{-1}(\Gamma/\Delta)^2$ of a scattering process ($\omega$ and $\omega_0$ are the laser and atomic angular frequencies, $\Gamma$ is the natural linewidth, $k=2\pi/\lambda$ is the wave number and $\overline{v}=\sqrt{k_BT/m}$ is a measure for the thermal speed of the atoms).
The left panel of Fig. \ref{Exp_scheme} shows a typical experimental CCD image, obtained at $T\simeq 195^\circ$C and $|\Delta|=2$ GHz 
\cite{footnote0, footnote1}. Under these conditions, $|\Delta|/k\overline{v}\simeq7.3\gg 1$ and $k\overline{v}/\Gamma\simeq 45\gg1$.
Despite the large Doppler effect, the image  in Fig. \ref{Exp_scheme} displays a well contrasted interference fringe, suggesting that atoms effectively behave like cold ones.
This fringe stems from the interference between counter-propagating wave paths scattered on both an atom and the mirror on the back face of the cell.
This process is illustrated in Fig. \ref{Exp_scheme}. It leads to an interference ring known as the ``mirror-assisted coherent backscattering'' (mCBS) effect, \cite{Greffet91, Labeyrie00, Moriya16} and on which we will concentrate our attention from now on.
While the mCBS of light was recently measured in a cold Strontium gas \cite{Moriya16}, its visibility in a \textit{hot} rubidium vapor with highly nontrivial quantum-level structure is far from obvious.
To understand it, we have analytically calculated the 
 enhancement factor $\Lambda\equiv(S_b+S_\text{mCBS})/S_b$ of the mCBS signal $S_\text{mCBS}$ with respect to the incoherent background signal $S_b$, taking into account the thermal distribution of atomic velocities:
\begin{align}
\label{EFtheta_eq}
\Lambda(\theta,T)=&1+\Lambda_0 
\exp\!\!\left[\!-2\bigg(\frac{k\overline{v}\theta_{\small{0}} L}{\Gamma\ell}\bigg)^{\!\!2}\right]\nonumber\\
&\times
\cos[kL (\theta-\theta_0)\theta_0]\text{sinc}[kL'(\theta-\theta_0)\theta_0],
\end{align}
where $\theta-\theta_0$ is the angular deviation from the fringe maximum, see Fig. \ref{Exp_scheme}, and $\Lambda_0\equiv \Lambda(\theta_0,T\!=\!0)-1$  
(the physical content of $\Lambda_0$  will be discussed later on).
Eq. (\ref{EFtheta_eq}) indeed describes an interference ring, with radial oscillations governed by two length scales, the distance $L/2$ from the center of the cell to the mirror, and the thickness $L'$ of the atomic vapor \cite{Moriya16}. The exponential factor stems from the thermal average of the dephasing $e^{i\Delta\Phi_{T}}$ accumulated by the two interfering paths, whose frequency is Doppler shifted by $\bs{v}\cdot\Delta{\bs{k}}$ upon scattering with momentum change $\Delta\bs{k}$ on an atom of velocity $\bs{v}$:
\begin{align}
\Delta\Phi_T\sim L {\bs{v}}\!\cdot\!\Delta{\bs{k}}\,\frac{\partial k}{\partial\omega} \sim \frac{Lk\overline{v}\theta_0}{\Gamma\ell},
\label{DeltaPhiT}
\end{align}
where we used that $\partial k/\partial \omega\sim 1/\Gamma\ell$ when $|\Delta|\gg\Gamma$, with $\ell$  the mean free path.
Eq. (\ref{DeltaPhiT}) explains the robustness of the mCBS interference against thermal motion. First, since $|\Delta|\gg k\overline{v}$ we have $\ell\simeq \ell(\overline{v}=0)\propto (\Delta/\Gamma)^2$, which lessens the impact of the thermal dephasing at large detuning. The second reason lies in the proportionality of $\Delta\Phi_T$ to the incident angle $\theta_0\ll1$: as we operate close to normal incidence, light scattering is essentially forward ($\Delta\bk\sim 0$) which again reduces $\Delta\Phi_T$.
Fig. \ref{mCBS_width} shows as blue dots a typical experimental  angular profile $\Lambda(\theta,T)$ of the mCBS ring, here measured at $|\Delta|=2$ GHz, for $\theta_0\simeq 7^\circ$ and for circularly polarized light detected in the channel of opposite helicity (channel $h\perp h$, see below). To obtain these data, we have measured both the background $S_b$ and mCBS $S_\text{mCBS}$ signals from which we have removed stray light by substracting the signal at $|\Delta|=40$ GHz.
The angular profile is compared with the theoretical prediction, Eq. (\ref{EFtheta_eq}), shown as a solid red curve. For this comparison, the amplitude at $\theta=\theta_0$, $\Lambda(\theta_0,T)$, is set to the experimental value 1.20. Except for this reference point though, there is no adjustable parameter: the agreement is excellent, in particular for the width of the central fringe and even for the first secondary fringes.

\begin{figure}[h]
\centering
\includegraphics[scale=0.78]{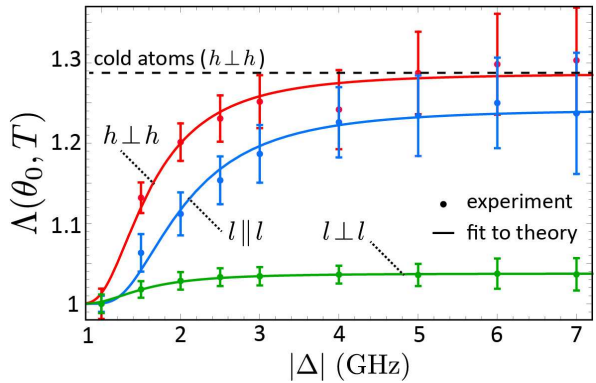}
\caption{(Color online) mCBS enhancement factor $\Lambda(\theta_0,T)$ as a function of detuning in three polarization channels. The  experimental data (dots) are fitted with Eq. (\ref{EFtheta_eq}) at $\theta=\theta_0$, using $\Lambda_0$ and $\rho k^{-3}$ as fit parameters (solid curves). These fits provide  $\rho k^{-3}=0.15$, 0.22, 0.14 and $\Lambda_0=0.038$, 0.242 and 0.287 in the channels $l\perp l$, $l\parallel l$ and $h\perp h$, respectively. No signal is observed in the channel $h\parallel h$. The dashed line shows the enhancement factor expected for cold atoms (i.e. in the limit $k\overline{v}/\Gamma\to 0$) in the channel $h\perp h$.
\label{EF_zerotheta}
}
\end{figure}

The most interesting property of Eq. (\ref{DeltaPhiT}) is the dependence of $\Delta\Phi_T$ on the mean free path $\ell\propto(\Delta/\Gamma)^2$. This offers the possibility to turn from a hot to a cold atom behavior in a controlled way via a change of the detuning. 
To check this property, we have  measured the mCBS enhancement factor $\Lambda(\theta=\theta_0,T)$ as a function of $|\Delta|$ over a broad range. The results are presented in Fig. \ref{EF_zerotheta}. 
The curves correspond to three detection schemes where linearly polarized light is analyzed along
the parallel $(l\parallel l)$ or perpendicular $(l\perp l)$ channels, or where circularly polarized
light is analyzed in the channel of 
opposite ($h\perp h$) helicity, as routinely done in experiments on light scattering in random media \cite{Labeyrie99, Wolf88}. No signal is observed in the channel $h\parallel h$ because at large detuning the excited hyperfine levels of the D2 line of rubidium are not resolved so that the single scattering process is essentially equivalent to a $J=1/2\to3/2$ transition \cite{Muller05}. Let us first discuss the variation of the curves with detuning.
We attribute it to the exponential factor in Eq. (\ref{EFtheta_eq}), which mainly governs the $\Delta$ dependence of  $\Lambda$. Fits of the experimental data to Eq. (\ref{EFtheta_eq}) (solid curves in Fig. \ref{EF_zerotheta}) validate this interpretation and demonstrate our ability to control the thermal dephasing of interfering wave paths via the detuning in the vapor. For comparison, in the channel $h\perp h$ we also show as a dashed line the enhancement  $\Lambda(\theta_0,T=0)=1+\Lambda_0$ expected for cold atoms. In fact, Fig. \ref{EF_zerotheta} indicates that the ``cold atom limit'' is almost reached for $\Delta\gtrsim$ 4GHz whatever the polarization configuration.

For the fits to Eq. (\ref{EFtheta_eq}), we use $\Lambda_0$ and the atom density $\rho k^{-3}$ as free parameters and extract in particular 
a nearly constant density $\rho k^{-3}=0.17\pm0.05$
, which confirms the independence of the dephasing (\ref{DeltaPhiT}) upon polarization. The relative values of enhancement factors in the various polarization channels thus  stem from the 
zero-temperature amplitude
$\Lambda_0=S_\text{mCBS}/S_b$. We attribute it 
 to the proportionality of the mCBS signal $S_\text{mCBS}(T\!=\!0)$ to the elastic atomic differential cross-section $d\sigma_\text{el}/d\Omega$, assuming that the background signal $S_b(T\!=\!0)$ is independent of polarization. Except at low detuning this assumption is approximately verified in our setup. For an incident wave field whose polarization vector changes from $\bs{\epsilon}_\text{in}$ to $\bs{\epsilon}_\text{out}$ upon scattering on an atom, it was shown, based on the decomposition of the scattered intensity into irreducible components with respect to the rotation group, that $d\sigma_\text{el}/d\Omega\propto w_1|\bs{\epsilon}_\text{in}\cdot \bs{\epsilon}_\text{out}^*|^2+w_2|\bs{\epsilon}_\text{in}\cdot \bs{\epsilon}_\text{out}|^2+w_3$ \cite{Omont77, Muller02}.
The three numerical coefficients $w_i$ depend on the fine and hyperfine level structure of rubidium. We have calculated them using the theoretical approach developed in \cite{Muller05} for treating light scattering from hyperfine multiplets. 
This calculation first confirms the absence of single scattering, and hence of mCBS fringe, in the channel $h\parallel h$ at large detuning.
In the other channels, it leads to the ratios
${\Lambda_0(h\perp h)}/{\Lambda_0(l\parallel l)}\simeq1.1$ 
and ${\Lambda_0(l\parallel l)}/{\Lambda_0(l\perp l)}\simeq8.0$ \cite{Cherroret18}, in reasonable agreement with the experimental ratios extracted from the fits in Fig. \ref{EF_zerotheta}: 
${\Lambda^{\text{exp}}_0(h\perp h)}/{\Lambda^{\text{exp}}_0(l\parallel l)}\simeq 1.2$ and ${\Lambda^{\text{exp}}_0(l\parallel l)}/{\Lambda^{\text{exp}}_0(l\perp l)}\simeq 6.4$.

\begin{figure}
\centering
\includegraphics[scale=0.7]{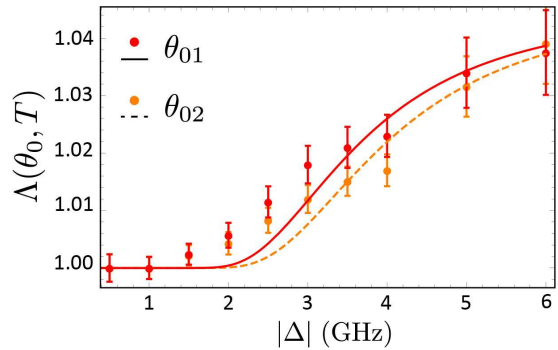}
\caption{(Color online) 
mCBS enhancement factor $\Lambda(\theta_0,T)$ versus detuning at two different incident  angles $\theta_0=\theta_{01}$ and $\theta_{02}>\theta_{01}$. Dots are experimental points and curves are fits  to Eq. (\ref{EFtheta_eq}). Here measurements are made in the channel $h\perp h$ at fixed $T\simeq168^\circ$. 
\label{EF_angle}
}
\end{figure}

According to Eq. (\ref{EFtheta_eq}), the thermal dephasing (\ref{DeltaPhiT}) can be also controlled with  the incident angle $\theta_0$. To verify this property, we have measured the mCBS enhancement factor as a function of detuning for two different angles $\theta_{01}\simeq6.5^\circ$ and $\theta_{02}\simeq7.5^\circ>\theta_{01}$ deduced from the mCBS angular profiles.
These measurements are displayed in  Fig. \ref{EF_angle} (dots). 
We fit them with Eq. (\ref{EFtheta_eq})  with $\Lambda_0$ as a single fit parameter (solid and dashed curves), inferring the atom density $\rho k^{-3}$ from the saturated vapor pressure in the cell.
For these measurements we used another atomic cell heated to $T\simeq168^\circ$ and on which no mirror was clipped on the back face (the glass itself thus plays the role of the mirror). $\Lambda_0$ being also proportional to the reflection coefficient of the glass \cite{Moriya16}, much smaller than the one of the mirror,
this leads to smaller enhancement factors than in Fig. \ref{EF_zerotheta}.

In the description of the mCBS interference presented so far, we implicitly assumed that light  was scattered \textit{elastically} on the atom (Rayleigh scattering). 
As we operate at a large detuning $|\Delta|\gg\Gamma$ though, the light-matter interaction may involve atomic transitions where several hyperfine levels come into play, such that it is not guaranteed that only Rayleigh scattering occurs. A quick look at the typical level structure of rubidium, recalled in Fig. \ref{inelastic_fringe}(b), confirms this statement:
\begin{figure}[h]
\centering
\includegraphics[scale=0.57]{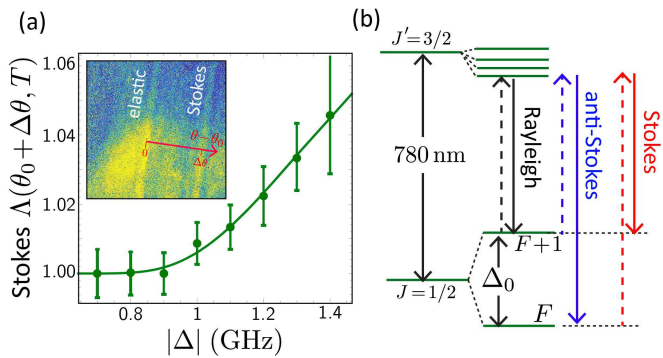}
\caption{(Color online) 
(a) Inset: CCD image showing the presence of a secondary mCBS ring associated with inelastic scattering, here visible in the channel $l\perp l$ at $|\Delta|=1.3$ GHz. This ring is angularly shifted with respect to the elastic one, as predicted by Eq. (\ref{EFtheta_eqin}) (the red arrow is the $\theta-\theta_0$ axis; the secondary ring is maximum at $\theta=\Delta\theta+\theta_0$).
Main panel: experimental  enhancement factor $\Lambda(\theta=\theta_0+\Delta\theta,T)$ of the secondary ring   as a function of detuning (dots), together with a fit to Eq. (\ref{EFtheta_eqin}).
(b) Level structure of rubidium. In addition to the usual elastic (Rayleigh) process, at large detuning light can experience inelastic (Stokes and anti-Stokes) processes from the two ground-state levels, separated by $\Delta_0$ [$\Delta_0=3.0$ GHz ($6.8$ GHz)  and $F=2$ ($F=1$) for $^{85}$Rb ($^{87}$Rb)].
\label{inelastic_fringe}
}
\end{figure}
the structure involves two ground-state levels $F$ and $F+1$, the various Zeeman sub-levels of which are equally populated in our hot vapor. When subjected to light, atoms may experience Rayleigh transitions $F\to F$ or $F+1\to F+1$ via any of the allowed excited levels [the transition $F+1\to F+1$ is illustrated in Fig. \ref{inelastic_fringe}(b)]. These processes, whose strength  is encapsulated in the elastic scattering cross-section, yield the elastic mCBS ring discussed previously. Since the detuning $\Delta$ is typically large compared to the spacing between the various excited hyperfine levels, those are not resolved in our experiment. $\Delta$ is, on the other hand, \textit{not} small compared to the spacing $\Delta_0$ between the two ground-state levels $F$ and $F+1$. This implies that Stokes ($F \to F+1$) and anti-Stokes ($F+1\to F$) \textit{inelastic} scattering processes can occur as well, see Fig. \ref{inelastic_fringe}(b). 
Note that this type of inelastic transitions here shows up in the \textit{linear} regime of light scattering (small saturation parameter) and is thus fundamentally different from the inelastic  scattering processes that occur at higher intensities \cite{Mollow69, Wellens04, Shatokhin05, Binninger18}. 
A refined analytical calculation of the mCBS effect shows that these inelastic processes give rise to two \textit{secondary} interference rings, of enhancement factor
\begin{align}
\label{EFtheta_eqin}
&\Lambda(\theta,T)=1+\Lambda_0 
\exp\!\!\left[\!-2\bigg(\frac{k\overline{v}\theta_{\small{0}} L}{\Gamma\ell}\bigg)^{\!\!2}\right]\\
&\times\cos[kL (\theta-\theta_0)\theta_0 +L\Delta k]\text{sinc}[kL'(\theta-\theta_0)\theta_0+L\Delta k]\nonumber.
\end{align}
As compared to Eq. (\ref{EFtheta_eq}), an additional momentum shift $\Delta k$ arises because the light frequency changes by $\pm\Delta_0$ after Stokes or anti-Stokes scattering on the atom. This shift  is given by
\begin{align}
\label{inelastic_kchange}
\Delta k=k\left[n(\Delta\pm\Delta_0)-n(\Delta)\right],
\end{align}
evaluated with a plus (minus) sign for the anti-Stokes (Stokes) process. In Eq. (\ref{inelastic_kchange}), $n$ is the refractive index of the atomic gas. The momentum shift in Eq. (\ref{EFtheta_eqin}) implies that inelastic rings are angularly separated by $\Delta\theta\simeq -\Delta k/(k\theta_0)$ from the elastic one. They are visible in polarization channels where inelastic scattering is present. We have found from calculations of inelastic scattering differential cross-sections that this is typically the case in the channel $l\perp l$, where we have experimentally focused our attention. In the inset of Fig. \ref{inelastic_fringe}(a) we show a  CCD image taken at $\Delta=-1.3$GHz and $T\simeq166^\circ$ in this channel: we indeed observe a secondary fringe close to the elastic one. 
To demonstrate that this signal is well of mCBS type, we have tested its sensitivity to thermal dephasing. 
The plot in Fig \ref{inelastic_fringe}(a) confirms this sensitivity: the detuning dependence of the enhancement factor $\Lambda(\theta_0+\Delta\theta,T)$ of the inelastic fringe  is well fitted by the exponential factor in Eq. (\ref{EFtheta_eqin}). From this fit we extract an atomic density $\rho k^{-3}\simeq 0.16$, which we use to estimate the angular separation $\Delta\theta$ between the elastic and the inelastic rings. 
Assuming a Stokes process (see below for the justification of this choice) and estimating the refractive index for a dilute atomic cloud, Eq. (\ref{inelastic_kchange})  with $\rho k^{-3}\simeq 0.16$ leads to $\Delta\theta=0.28^\circ$ \cite{footnote2}. This value is on the order of the  experimental one, $\Delta\theta_\text{exp}=0.15^\circ$, measured on the camera image. A possible reason for the theoretical overestimation is the uncertainty on the atom density $\rho k^{-3}$. 
 One may wonder, finally, why  only one inelastic fringe is visible in the inset of Fig. \ref{inelastic_fringe}, while Eq. (\ref{inelastic_kchange}) in principle predicts two fringes. The reason lies in the frequency asymmetry of the Stokes and anti-Stokes transitions. Indeed, unlike the Stokes process, the anti-Stokes process brings photons back to resonance \cite{footnote1}. This  leads to a large momentum shift (\ref{inelastic_kchange}), which moves the anti-Stokes ring far away from the elastic one, out of the range of the camera [from Eq. (\ref{inelastic_kchange}) we estimate $\Delta\theta_{\text{anti-Stokes}}\sim -10 \Delta\theta_{\text{Stokes}}$). Note that the Stokes nature of the secondary fringe in Fig \ref{inelastic_fringe}(a) is also confirmed by its position with respect to the elastic one: for the Stokes process $n(\Delta-\Delta_0)-n(\Delta)<0$ so that $\Delta \theta>0$.
 
We have shown that coherent transport of light in hot atomic vapors can survive large Doppler effects. 
This paves the way to the future observation of other mesoscopic phenomena in hot vapors, like coherent backscattering, which can be enhanced by operating in the large detuning limit or with a magnetic field \cite{Sigwarth04}. We have also 
experimentally and theoretically unveiled novel interference effects emerging from the internal multi-level structure. This sheds light on the important role inelastic Raman processes might play in multiple scattering of light \cite{Priolo17}, a question yet largely  unexplored  in ensembles of quantum scatterers. 

NC thanks Dominique Delande for helpful discussions, and financial support from and the Agence Nationale de la Recherche (grant ANR-14-CE26-0032 LOVE). This work was conducted within the framework of the project OPTIMAL granted by the European Union by means of the Fond Europ\'een de d\'eveloppement r\'egional, FEDER.


\begin{thebibliography}{99}

\bibitem{Akkermans07}
E. Akkermans and G. Montambaux, 
\textit{Mesoscopic physics of
electrons and photons} (Cambridge university press, 2007).

\bibitem{Hulst12}
H. C. Van de Hulst, \textit{Multiple light scattering: tables, formulas,
and applications} (Elsevier, 2012).


\bibitem{Tiggelen08}
H. Hu, A. Strybulevych, J. Page, S. Skipetrov, and B. van
Tiggelen, Nat. Phys. \textbf{4}, 945 (2008).

\bibitem{Gigan17}
S. Rotter and S. Gigan,
Rev. Mod. Phys. \textbf{89}, 015005 (2017).

\bibitem{Molisch1998}
A. F.  Molisch and B. P.  Oehry, 
\textit{Radiation trapping in
atomic vapors, Clarendon Press}, Oxford (1998).

\bibitem{Aspect09}
A. Aspect and M. Inguscio, Physics today \textbf{62}, 30 (2009).





\bibitem{Dalibard1989}
J. Dalibard and C. Cohen-Tannoudji,
J.O.S.A B \textbf{11}, 2043 (1989).

\bibitem{Hau1999}
L. V. Hau, S. E. Harris, Z. Dutton, and C. H. Behroozi,
Nature, \textbf{397}, 6720 (1999).

\bibitem{DLCZ}
L. M. Duan,  M. D. Lukin,  J. I. Cirac, and P. Zoller,
Nature \textbf{414}, 413 (2001).



\bibitem{Guerin2016}
W. Gu\'erin, M. O. Ara\`ujo, and R. Kaiser,
Phys. Rev. Lett. \textbf{116}, 083601 (2016).

\bibitem{Pellegrino14}
J. Pellegrino, R. Bourgain, S. Jennewein, Y. R. P. Sortais, A. Browaeys, S. D. Jenkins, and J. Ruostekoski,
Phys. Rev. Lett. \textbf{113}, 133602 (2014).

\bibitem{Browaeys2016}
S. Jennewein, M. Besbes, N. J. Schilder, S. D. Jenkins, C. Sauvan, J. Ruostekoski, J.-J. Greffet, Y. R. P. Sortais, and A. Browaeys, 
Phys. Rev. Lett. \textbf{116}, 233601 (2016).

\bibitem{Beugnon2017}
 L. Corman, J. L. Ville, R. Saint-Jalm, M. Aidelsburger, T. Bienaim\'e, S. Nascimb\`ene, J. Dalibard, and J. Beugnon, 
Phys. Rev. A \textbf{96}, 053629 (2017).

\bibitem{Saint-Jalm2018}
R. Saint-Jalm, M. Aidelsburger, J. L. Ville, L. Corman, Z. Hadzibabic, D. Delande, S. Nascimb\`ene, N. Cherroret, J. Dalibard, and J. Beugnon,
Phys. Rev. A \textbf{97}, 061801(R) (2018).

\bibitem{Stenflo2015}
J. O. Stenflo,
ApJ \textbf{801}, 70 (2015).

\bibitem{Shaffer2013}
J. Sedlacek, A. Schwettmann, H. K\"ubler,  and J. Shaffer,
Phys. Rev. Lett. \textbf{111}, 063001 (2013).

\bibitem{Pfau2015}
A. Urvoy, F. Ripka, I. Lesanovsky, D. Booth, J. P. Shaffer, T. Pfau and R. L\"ow, 
Phys. Rev. Lett. \textbf{114}, 203002 (2015).

\bibitem{Hughes2017}
D. J. Whiting, N. Sibalic, J. Keaveney, C. S. Adams, and I. G. Hughes, 
Phys. Rev. Lett. \textbf{118}, 253601 (2017).

\bibitem{Polzik2017}
C. B. Moller, R. A.Thomas, G. Vasilakis, E. Zeuthen, Y. Tsaturyan, M. Balabas, K. Jensen, A. Schliesser, K. Hammerer, and E. S. Polzik, 
Nature \textbf{547}, 191 (2017).

\bibitem{Mitchell2018}
R. Jim\'enez-Mart\'inez,  J.Kolodynski,  C. Troullinou,  V. G. Lucivero,  J. Kong,  and M. W. Mitchell, 
Phys.  Rev.  Lett. \textbf{120}, 040503 (2018).

\bibitem{Sheng95}
P. Sheng, \textit{Introduction to Wave Scattering, Localization, and Mesoscopic Phenomena, Academic Press, New York, (1995). }


\bibitem{Labeyrie99}
G. Labeyrie, F. de Tomasi, J.-C. Bernard, C. A. M\"uller, Ch. Miniatura and R. Kaiser,
Phys. Rev. Lett. \textbf{83}, 5266 (1999).

\bibitem{Golubentsev84}
A. A. Golubentsev, Sov. Phys. JETP \textbf{59}, 26 (1984).

\bibitem{Labeyrie06}
G. Labeyrie, D. Delande, R. Kaiser, and C. Miniatura,
Phys. Rev. Lett. \textbf{97}, 013004 (2006).

\bibitem{footnote0}
Here and in the rest of the Letter, temperatures are given with an uncertainty of about $10^\circ$ C.

\bibitem{footnote1}
For convenience, in the experiment we choose to work with negative detunings.

\bibitem{Greffet91}
J.-J. Greffet, Waves Random Media \textbf{3}, S65 (1991).

\bibitem{Labeyrie00}
G. Labeyrie, C. A. M\"uller, D. S. Wiersma, C. Miniatura, and R Kaiser, 
J. Opt. B: Quant. Semiclass. Opt. \textbf{2}, 672 (2000).

\bibitem{Moriya16}
P. H. Moriya, R. F. Shiozaki, R. Celistrino Teixeira, C. E. M\'aximo, N. Piovella, R. Bachelard, R. Kaiser, and Ph. W. Courteille,
Phys. Rev. A \textbf{94}, 053806 (2016).

\bibitem{Wolf88}
P. E. Wolf, G. Maret, E. Akkermans and R. Maynard, 
J. Phys. (France) \textbf{49}, 63 (1988).

\bibitem{Muller05}
C. A. M\"uller, C. Miniatura, D. Wilkowski, R. Kaiser, and D. Delande,
Phys. Rev. A \textbf{72}, 053405 (2005).


\bibitem{Omont77}
A. Omont, 
Prog. Quantum Electron. \textbf{5}, 69 (1977).

\bibitem{Muller02}
C. A. M\"uller and C. Miniatura, 
J. Phys. A \textbf{35}, 10163 (2002).

\bibitem{Cherroret18}
N. Cherroret \textit{et al.}, to be published.

\bibitem{Mollow69}
B. R. Mollow, Phys. Rev. \textbf{188}, 1969 (1969).

\bibitem{Wellens04}
T. Wellens, B. Gr\'emaud, D. Delande, and C. Miniatura, 
Phys. Rev. A \textbf{70}, 023817 (2004).

\bibitem{Shatokhin05}
V. Shatokhin, C. A. M\"uller, and A. Buchleitner, 
Phys. Rev. Lett. \textbf{94}, 043603 (2005).

\bibitem{Binninger18}
T. Binninger, V. N. Shatokhin, A. Buchleitner, and T. Wellens,
arXiv:1811.08882 (2018).


\bibitem{footnote2}
This value corresponds to the isotope $^{85}$Rb for which $\Delta_0=3.0$ GHz. The isotope $^{87}$Rb, for which $\Delta_0=6.8$ GHz, in principle gives rise to an extra Stokes ring but with larger separation $\Delta\theta\simeq 0.37^\circ$.






\bibitem{Sigwarth04}
O. Sigwarth, G. Labeyrie, T. Jonckheere, D. Delande, R. Kaiser, and C. Miniatura,
Phys. Rev. Lett. \textbf{93}, 143906 (2004).

\bibitem{Priolo17}
B. Fazio, A. Irrera, S. Pirotta, C. D'Andrea, S. Del Sorbo, M. J. Lo Faro, P. G. Gucciardi, M. A. Iat\`i, R. Saija, M. Patrini, P. Musumeci, C. S. Vasi, D. S. Wiersma, M. Galli, and F. Priolo,
Nat. Phot. \textbf{11}, 170 (2017).




















 












\end{thebibliography}
\end{document}